\documentclass[conference]{IEEEtran}
\ifCLASSINFOpdf
\usepackage[pdftex]{graphicx}
\else
\fi
\usepackage{color}
\usepackage{float}

\usepackage{amsmath}
\usepackage{algorithmic}

\usepackage{array}
\begin{document}
%
\title{Estimating Emotional Intensity from Body Poses for Human-Robot Interaction}

\author{
\IEEEauthorblockN{Mingfei Sun} 
\IEEEauthorblockA{Computer Science \& Engineering\\
Hong Kong University of Science and Technology\\
mingfei.sun@ust.hk}
\and
\IEEEauthorblockN{Yiqing Mou}
\IEEEauthorblockA{Psychology Department\\
Hong Kong University\\
myq.maxine@gmail.com}
\and
\IEEEauthorblockN{Hongwen Xie}
\IEEEauthorblockA{Tencent Inc.\\
hongwenxie@tencent.com}
\and
\IEEEauthorblockN{Meng Xia}
\IEEEauthorblockA{Computer Science \& Engineering\\
Hong Kong University of Science and Technology \\
iris.xia@connect.ust.hk}
\and
\IEEEauthorblockN{Michelle Wong}
\IEEEauthorblockA{Computer Science\\
Smith College \\
mwong46@smith.edu}
\and
\IEEEauthorblockN{Xiaojuan Ma}
\IEEEauthorblockA{Computer Science \& Engineering\\
Hong Kong University of Science and Technology \\
mxj@cse.ust.hk}
}


%
%


\maketitle

\begin{abstract}
Equipping social and service robots with the ability to perceive human emotional intensities during an interaction is in increasing demand. Most of existing work focuses on determining which emotion(s) participants are expressing from facial expressions but largely overlooks the emotional intensities spontaneously revealed by other social cues, especially body languages. In this paper, we present a real-time method for robots to capture fluctuations of participants' emotional intensities from their body poses. Unlike conventional joint-position-based approaches, our method adopts local joint transformations as pose descriptors which are invariant to subject body differences as well as the pose sensor positions. In addition, we use a Long Short-Term Memory Recurrent Neural Network (LSTM-RNN) architecture to take the specific emotion context into account when estimating emotional intensities from body poses. The dataset evaluation suggests that the proposed method is effective and performs better than baseline method on the test dataset. Also, a series of succeeding field tests on a physical robot demonstrates that the proposed method effectively estimates subjects emotional intensities in real-time. Furthermore, the robot equipped with our method is perceived to be more emotion-sensitive and more emotionally intelligent.

\end{abstract}


%
\IEEEpeerreviewmaketitle

\section{Introduction}

\begin{figure}
\centering
\includegraphics[width=0.8\linewidth]{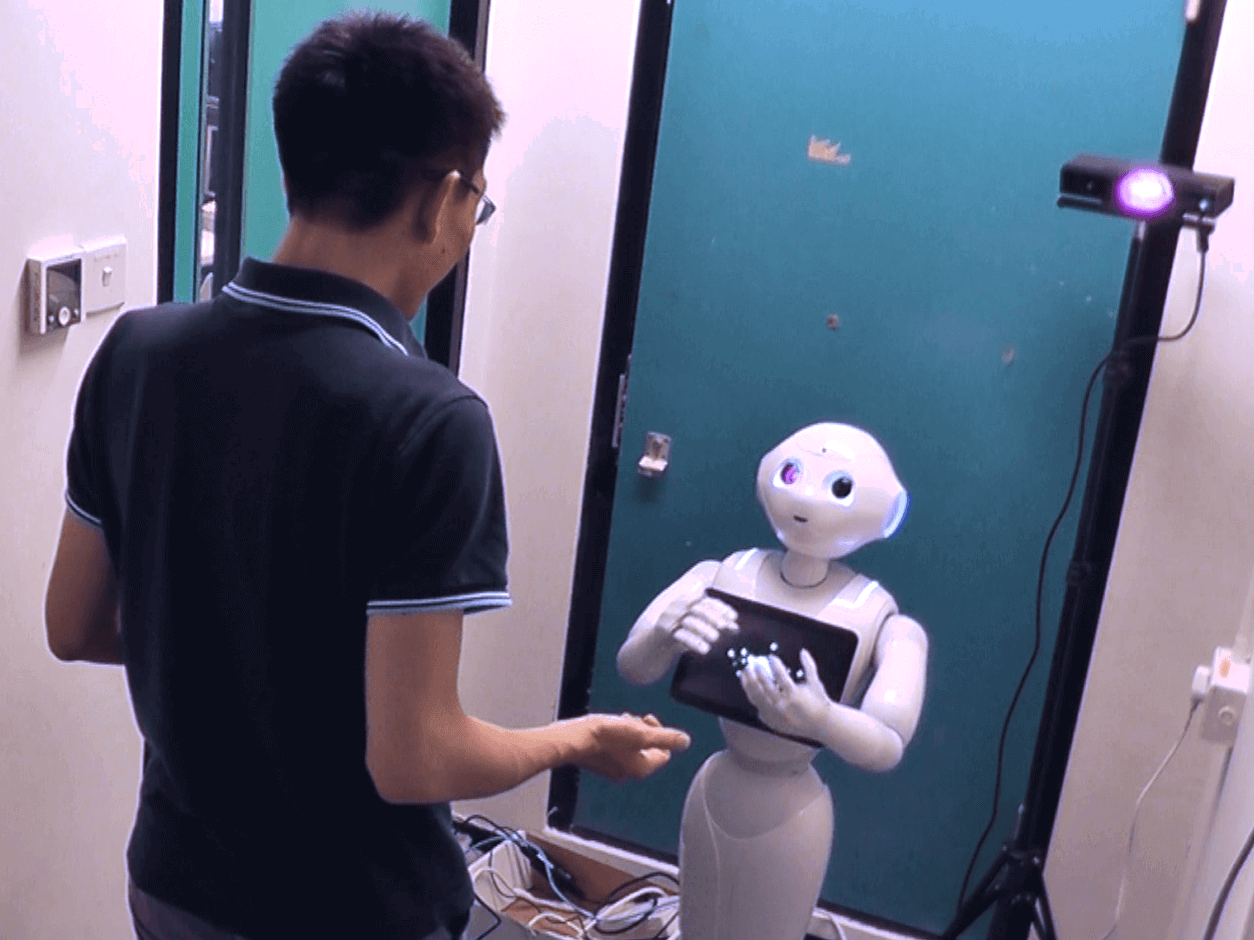}
\caption{A subject shows his high intensity of surprise by spreading out his hands.}
\label{fig:main}
\end{figure}

There is an increasing demand in Human-Robot Interaction (HRI) for real-time perception of participants' emotional intensities \cite{gonsior2012emotional}, i.e., how emotions fluctuate over time. Human participants may express their feelings and intents through subtle emotional intensities in HRI \cite{mccoll2016survey}. For example, when users become unsatisfied with robots' performance, they may frown. If agitated by robots' inappropriate responses, they may shrug to show their strong disappointments. If the robot is capable of detecting such fluctuations of emotional intensities, it can then fine-tune its reactions in a timely manner to lower participants' discomfort \cite{rani2006affective}, to satisfy their preferences \cite{liu2008online}, and consequently to gain more acceptance \cite{sorbello2014telenoid}. 
In spite of such demand, most of existing work focuses on determining which emotion(s) participants are expressing from facial expressions \cite{cid2013real} but largely overlooks the emotional intensities spontaneously revealed via other social cues, especially via body languages. Particularly,  
when emotions get more intense, the discriminative power of facial expressions for emotional intensities will degrade whereas the body cues will become dominant \cite{aviezer2012body}.
Nevertheless, a dominant increase of body poses is not necessarily an indicator of more emotional intensity, since the latter is usually emotion-specific \cite{wallbott1998bodily}. Take Fig.~\ref{fig:main} as an example. Open arms may just be an ordinary gesture if the person is speaking calmly. But when he is with a surprised look, the pose may imply more surprise than that when his arms are in resting positions. Hence, in order to get a more complete picture of the emotional intensity from body poses, it is better to integrate emotion types into the intensity estimation process. Furthermore, most of previous studies on estimating affective status from body poses directly take absolute joint positions \cite{gunes2007bi} or other physical quantities derived from these positions, e.g., joint speeds \cite{xu2014robot}, body expansions \cite{mccoll2014determining} etc., as pose descriptors. Regardless of the reported effectiveness, these features, in general, are subject to individuals' body differences (such as heights, body shapes etc.) as well as their positions with respect to pose sensors, which may possibly limit or even deprive their practical usages.

In this paper, we present a novel method to estimate emotional intensities from body poses under various affective states. Specifically, we adopt the local joint transformations to describe body poses, which are invariant to subjects' body differences as well as their positions against sensors. Furthermore, instead of deriving handcrafted features from joint transformations, we use a Long Short-Term Memory Recurrent Neural Network (LSTM-RNN) to directly model the pose sequences with varied lengths. This has two advantages. First, the LSTM-RNN captures complex temporal correlations among pose sequences and thus provides a more appropriate modeling of fluctuations of emotional intensities. Second, through proper design of neural network structures, we integrate emotional contexts into intensity estimation, and obtain more accurate predications of emotional intensities. 
Quantitative evaluations on a dataset demonstrate that the proposed method accurately predicts emotional intensities with an average Pearson correlation score of around 0.8, outperforming the previous SVM-based method which achieves only 0.46. We also deploy the proposed method to a physical robot (Pepper robot) and conduct a series of field tests to evaluate its practical usage. The results imply that the robot equipped with our method senses subjects' emotional intensities effectively in real-time. Also, the subjects' ratings suggest that the robot with the proposed emotional intensity estimation is perceived to be more emotional-sensitive and emotionally intelligent.

\section{Related Work}
Emotional intensity is defined as the psychological states of being activated by some evoking stimulus \cite{russell1999core}, which can be measured by bio-electric signals, such as heart rate, blood pressure, skin conductance etc \cite{khalfa2002event,appelhans2006heart,james1986influence}. For example, Kulic and Croft leveraged heart rate, perspiration rate, and facial muscle contraction to estimate participants' emotional intensities in terms of anxiety, calm and surprise towards robot arm motions \cite{kulic2007affective}; Swangnetr and Kaber used heart rate and galvanic skin response signals to estimate the intensity of joy and excitement for the elderly patients during medical Patient-Robot Interactions \cite{swangnetr2013emotional}. Saulnier \emph{et al.} exploited the natural muscle tensions to estimate participants' stress levels for domestic robots \cite{saulnier2009using}. However, all these studies require putting sensor on participants, e.g., via special wearable devices \cite{rani2006affective} or electro-physiological monitoring devices \cite{kulic2007affective,liu2008online}, it may be difficult and sometimes impossible to deploy such biological methods for social or service robots in real-time HRI.

Non-contact affective intensity sensing is thus proposed, which usually adopts different modalities to estimate the emotional intensity. Among all modalities, the body pose has gained increasing popularity in HRI due to the following two reasons. First, body poses are becoming readily accessible in HRI due to the real-time motion capture sensors and robust vision estimation techniques; Second, more and more psychological results demonstrate the strong possibility of using body poses to perceive emotions \cite{wallbott1986cues,dittrich1996perception,wallbott1998bodily,pollick2001perceiving}. Although whether body poses are indicative of specific emotions is still debatable \cite{wallbott1998bodily}, psychologists have reached a consensus that body poses can effectively convey people's emotional intensity \cite{ekman1974detecting}. Thus many researchers employ body poses as an important indicator of participants' emotional intensity. For example, McColl and Nejat designed a model to recognize the elderly people's pleasure intensity based on their upper body movements during a dining process with a robot assistant \cite{mccoll2014determining}. Their results indicated that, compared with the specific emotion types, the emotional intensities can be  more accurately perceived from body movements. In addition, Xu \emph{et al.} employed an expressive Nao robot to emotionally influence participants' body poses \cite{xu2014robot}. Their results also indicated that the emotional arousal can be effectively conveyed via body poses. Other studies have also identified the strong link between a subject's emotional intensities and the body movements \cite{kleinsmith2007recognizing} as well as the head positions \cite{gross2012effort}. Furthermore, the body poses are also adopted to perceive other affective and cognitive process, e.g., accessibility level \cite{mccoll2012affect}, engagement dynamics \cite{sanghvi2011automatic,sun2017sensing}.  However most of the aforementioned works adopt the absolute joint positions or handcrafted features, which could result in two problems. First, the features can be affected by irrelevant physical body differences. For example, joint speeds\cite{xu2014robot}, joint accelerations \cite{saha2014study}, joint distances \cite{mccoll2014determining,kleinsmith2007recognizing}, body expansion\cite{mccoll2014determining} and silhouette motion images \cite{castellano2007recognising} are closely related to subjects' heights and body shape (e.g., bone lengths). Thus, for subjects with different heights and body shapes, the corresponding features will have distinguishable differences even if subjects perform the same set of expressive body motions. In addition, the joint-position-based features is also dependent on camera positions. For example, the joint angles, and the joint structures will change dramatically if the pose sensors are relocated to another position. Furthermore, as pointed out in \cite{wallbott1998bodily}, the emotional intensity expressed via body poses are specific to emotion types, which is also largely overlooked by previous studies. 

\section{Method}

In this section, we describe the proposed method in detail.

\subsection{Body pose representation}
\begin{figure}
\centering
\includegraphics[width=1.0\linewidth]{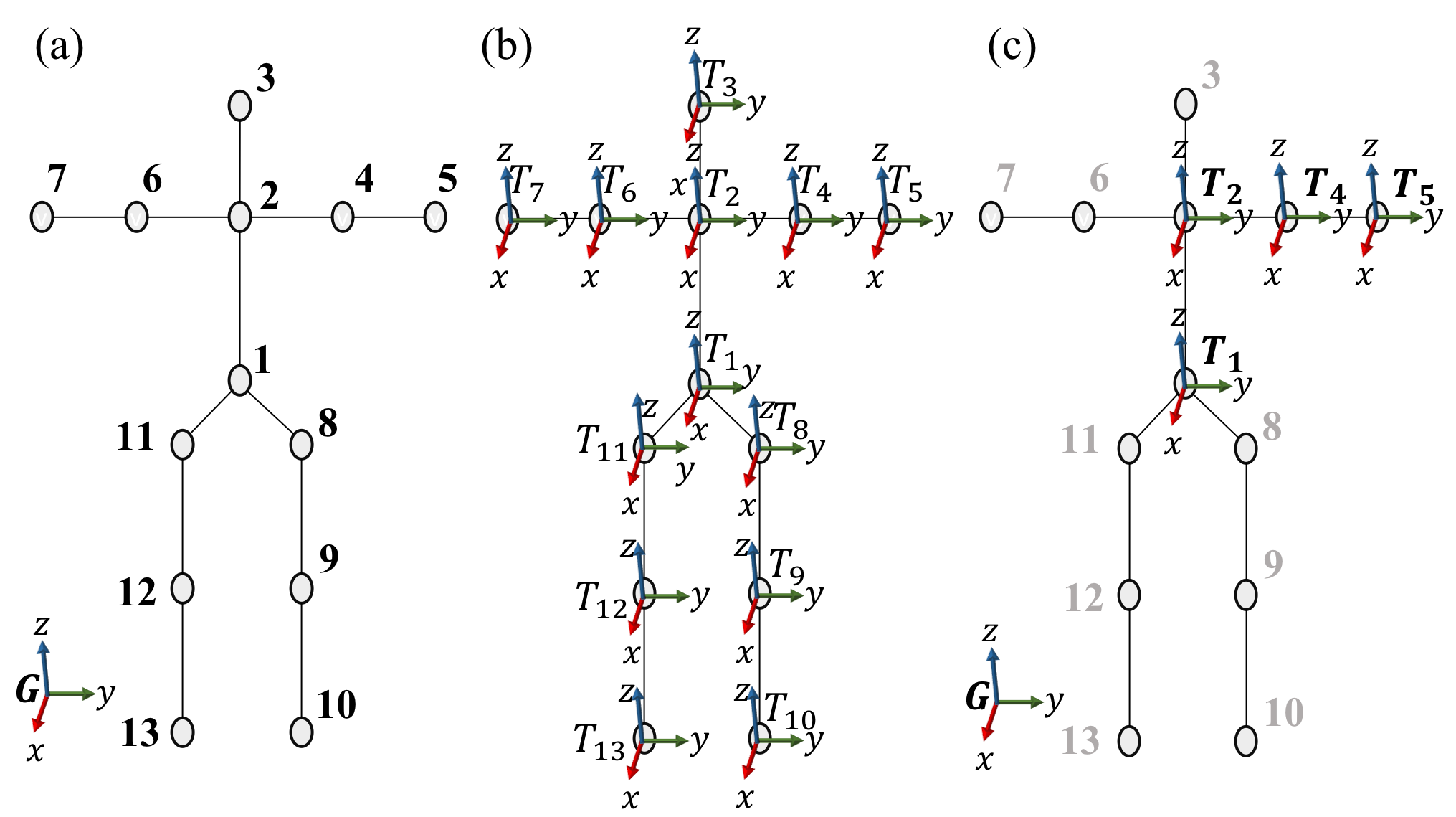}
\caption{(a) Skeletal tree structure with the joint order number; (b) Each joint has its own coordination frame and all frames are ordered; (c) To find the transformation $T_5$ for joint$\#5$ with position $G_5$, we first compute forward kinematics along transformation tree: $G_5 = T_1 T_2 T_4 T_5$, and then $T_5 = T_4^{-1} T_2^{-1} T_1^{-1} G_5 $ }
\label{fig:pose}
\end{figure}

The body pose is usually denoted as a skeletal tree with parent-child structures, as shown in Fig.~\ref{fig:pose}(a). Each node in the tree is associated with a body joint, and its coordinates represent the joint position with respect to the sensor coordinate frame. This representation is widely adopted for affective estimation in most of the existing literatures. However, as aforementioned, the body pose represented in this way is dependent on the sensor positions and contains individual body differences, e.g., bone lengths(the distance between two joints), which could limit the generalization of the proposed methods in previous work.

Nevertheless, human body poses can also be described via a series of local joint transformations, as widely used in motion capture\cite{moeslund2006survey} and motion re-targeting\cite{gleicher1998retargetting}. The basic idea is to separate relative joint displacements and motions in skeletal structures by using local transformations in joint coordinate frames. Specifically, each joint has its own coordination frame (a predefined right-handed frame) and all frames are ordered based on the skeletal tree structure, forming multiple parent-child pairs, as shown in Fig.~\ref{fig:pose}(b). The position of each joint is defined by a homogeneous matrix, which contains the joint rotation and translation with respect to its parental coordinate frame. This transformation representation describes how joints move against their parental joints, i.e., the joint relative motions rather than absolute positions, thus invariant to sensor positions. To get rid of subject-dependent body differences, we discard translations in the homogeneous matrix and only adopt rotations since joint translations are relevant to bone length. For simplicity, we use Euler angles $(\theta_{r}, \theta_{p}, \theta_{y})$ in the order of roll, pitch and yaw to represent rotations in a homogeneous matrix. Consequently, each body pose is described by a vector of Euler angles, ordered according to the skeletal tree structure. In addition, we do not consider rotations of the root joint as it is in the sensor coordinate frame. For each body pose $X_i$ in a pose sequence, we thus have following descriptions:
\begin{equation*}
X_i = \begin{bmatrix}
    \theta_{r1} & \theta_{r2} & \theta_{r3} & \dots  & \theta_{r13} \\
    \theta_{p1} & \theta_{p2} & \theta_{p3} & \dots  & \theta_{p13} \\
    \theta_{y1} & \theta_{y2} & \theta_{y3} & \dots  & \theta_{y13}
\end{bmatrix}
\end{equation*}
where $\theta \in [-\pi, \pi]$.

In most of existing pose extracting sensors, e.g., Kinect, the local joint transformation representation is readily available. Nevertheless, we can also easily convert the skeletal tree representation into the local transformations. The conversion between these two representations is essentially a set of transformations between different frames, which are usually described by forward/inverse kinematics in robotics. As illustrated in the example of Fig.~\ref{fig:pose}(c), the conversion from local joint transformation $T_{i}$ of joint$\#i$ to its positions $G_i$ in skeletal tree can be found by:
\begin{equation*}
G_i = T_1 T_{i_2} T_{i_3} ... T_{i}
\end{equation*}
where $G_i \in \mathbf{R}^{4}$ is the homogeneous coordinate of joint$\#i$ in the sensor coordinate frame, $T_i$ is the transformation matrix (homogeneous matrix) of joint$\#i$ in its parental frame, and $i, i_{1}, i_{2}, ..., 1$ is a path from joint$\#i$ to join$\#1$. To find the local transformation representation for joint$\#i$, we can use:
\begin{equation*}
T_i = T_{i}^{-1} T_{i_1}^{-1} T_{i_2}^{-1} ... T_1^{-1} G_i
\end{equation*}
In addition,

\subsection{LSTM-RNN modeling}
\begin{figure}
\centering
\includegraphics[width=1.0\linewidth]{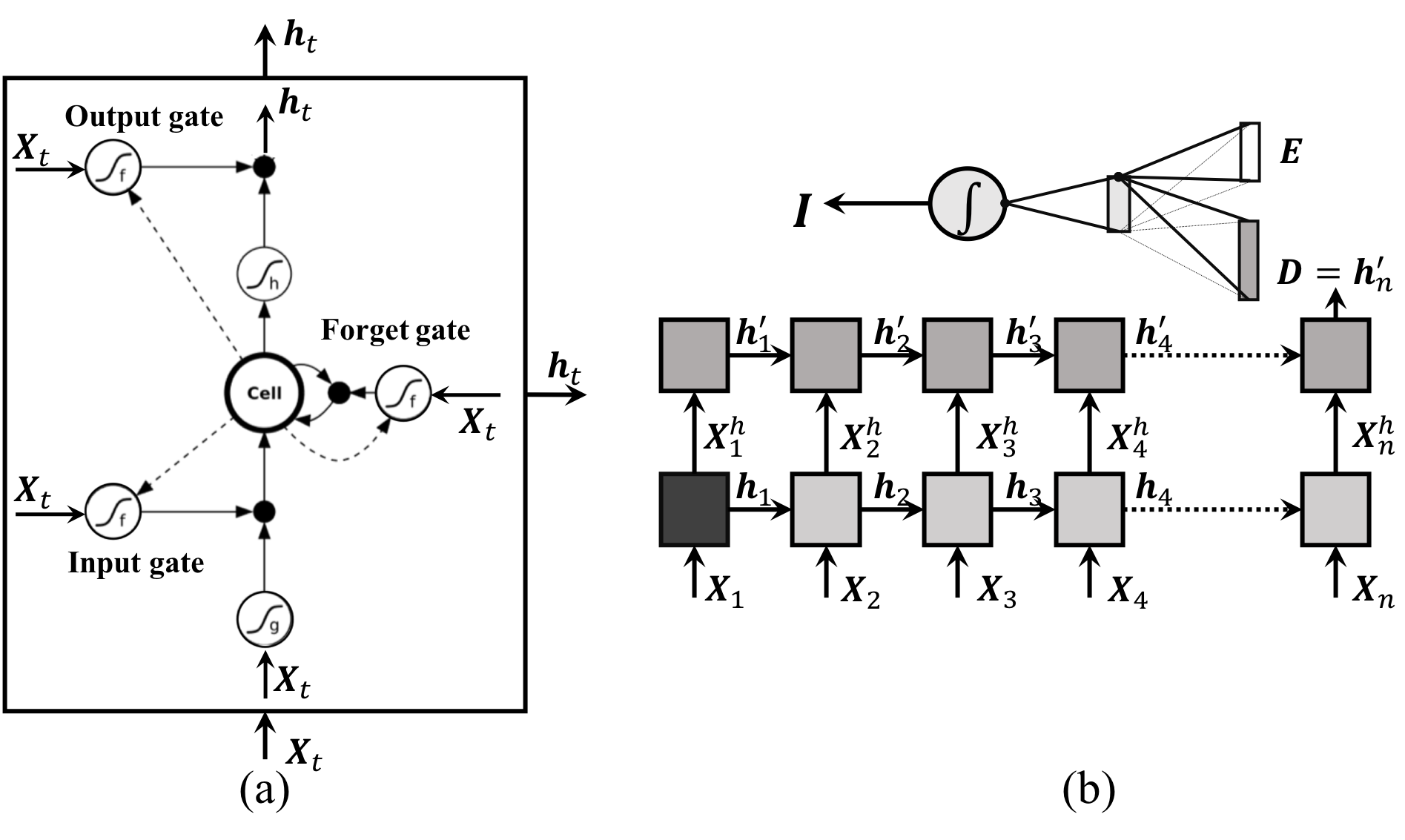}
\caption{(a) LSTM memory block; (b) the proposed LSTM architecture}
\label{fig:lstm}
\end{figure}

Owing to its capability of modeling complex temporal dependencies, Long Short-Term Memory Recurrent Neural Network (LSTM-RNN) has been widely adopted for pose-related estimation, e.g., human pose estimation\cite{liu2016spatio}, action recognition\cite{srivastava2015unsupervised} etc. Fig.~\ref{fig:lstm}(a) presents the internal structure of an LSTM memory block, which has a single cell and three multiplicative gates (input gate, forget gate and output gate). The detailed multiplicative relationships between these gates can be referred to \cite{graves2012supervised}. LSTM-RNN can capture temporal flows along thousands or even millions of time steps in sequences with variable lengths, which is usually hard to achieve for conventional sequence models, e.g., Hidden Markov Model. These advantages make LSTM-RNN a good choice for processing the complicated, variable-length and intra-correlated body pose sequences.

Inspired by the work \cite{donahue2015long} that uses a two-layered LSTM-RNN to model a sequence of structured features, we adopt an LSTM-RNN architecture as shown in Fig.~\ref{fig:lstm}(b). It is composed of two LSTM layers and one fully-connected layers. The LSTM layers take pose sequences $[X_1, X_2, X_3, ..., X_n]$ of variable lengths $n$ as input and output a pose description vector $D=h_{n}^\prime$ with fixed size. The fully-connected layer fuses the emotion vector $E$ (emotional context of body poses) with pose descriptor $D$ and outputs a estimated scalar $I$ of emotional intensity by a sigmoid function.

\section{Experiments \& Results}
In order to evaluate our system, we conduct two studies: empirical evaluation on a public dataset and a field test on a physical robot. 

\subsection{Dataset evaluation}
We use the body expression dataset\footnote{Provided by the Max-Planck Institute for Biological Cybernetics; downloadable: http://ebmdb.tuebingen.mpg.de/} for the evaluation of the proposed method. This dataset contains close-to-natural emotional body expressions recorded by motion capture devices when subjects were expressively narrating stories in front of the camera. Each pose is described by $23$ body joints in 3D with recording rate at $120Hz$, and each body joint$\#i$ has a local transformation $T_i$. The dataset has two labels: the intended emotion and the perceived emotion. The intended emotion is what the narrator is required to express, i.e., $E$ in the input of LSTM-RNN, while the perceived emotion is a set of emotion labels annotated by experts. We use the percentage of the intended emotion in the set of the perceived emotion as the ground truth for emotional intensity $I$ based on the intuition that: if the emotion expressed via a pose sequence is intenser, then annotators are more likely to assign the same and correct label to it. The dataset contains 1447 sequences of natural emotional body poses, and each sequence is stored in local transformation format.

Since the performance of a neural network depends heavily on the size of training dataset, we augment the training data to obtain more data in two steps. First, we double the dataset by swapping left and right side of body poses since this swap will not change the expressed emotional intensity. Second, we sample from raw pose sequences (120Hz) at a fixed rate(30Hz) and get four samples from every pose sequence, then further quadrupling the dataset. Finally, we have around 12,000 body sequences.

The LSTM RNN is trained with following setups: $h \in \mathbf{R}^{64}$, $h^{\prime} \in \mathbf{R}^{128}$ for hidden units; 100 epochs with batch size 256 for one complete training. We use RMSprop as the optimizer with mean absolute error as the loss.

To show the effectiveness of the proposed method, we use the SVM together with handcrafted features proposed in \cite{kapur2005gesture,piana2016adaptive} as the baseline, which adopts joint speeds, joint accelerations, joint angles and body expansions to estimate emotion from postures. The Pearson correlation coefficient\footnote{The definition can be found in: https://en.wikipedia.org/wiki/ Pearson\_correlation\_coefficient} is used as the evaluation metric. We use 5-fold cross-validation to compare the results. 


\begin{table}
\centering
\includegraphics[width=1.0\linewidth]{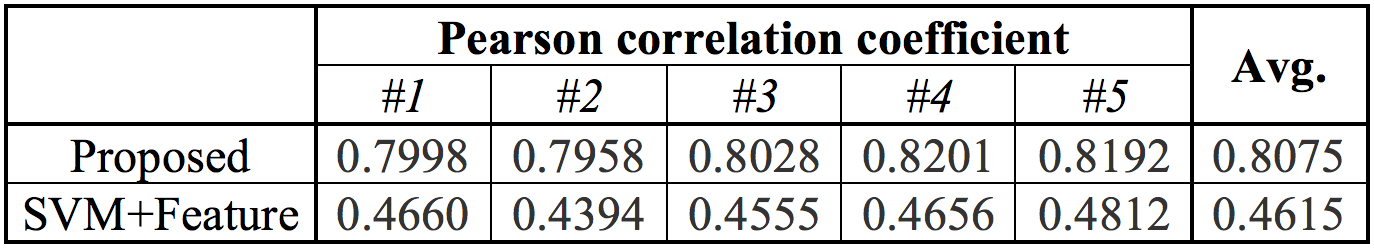}
\caption{Five-fold cross validations on the MPI dataset}
\label{tab:cv-test}
\end{table}

The results in Tab.\ref{tab:cv-test} demonstrate that the proposed method effectively estimates the emotional intensities. Compared with the baseline method (SVM with handcrafted features), our method consistently achieves higher Pearson correlation coefficients on all tests. In addition, the average coefficient of the proposed method is almost 75\% higher than the baseline, which further confirms that the adopted pose descriptors are superior to the joint-position-based handcrafted features. And LSTM-RNN is also effective for intensities estimation from body poses.

\subsection{Field testing}
\begin{figure}
\centering
\includegraphics[width=0.9\linewidth]{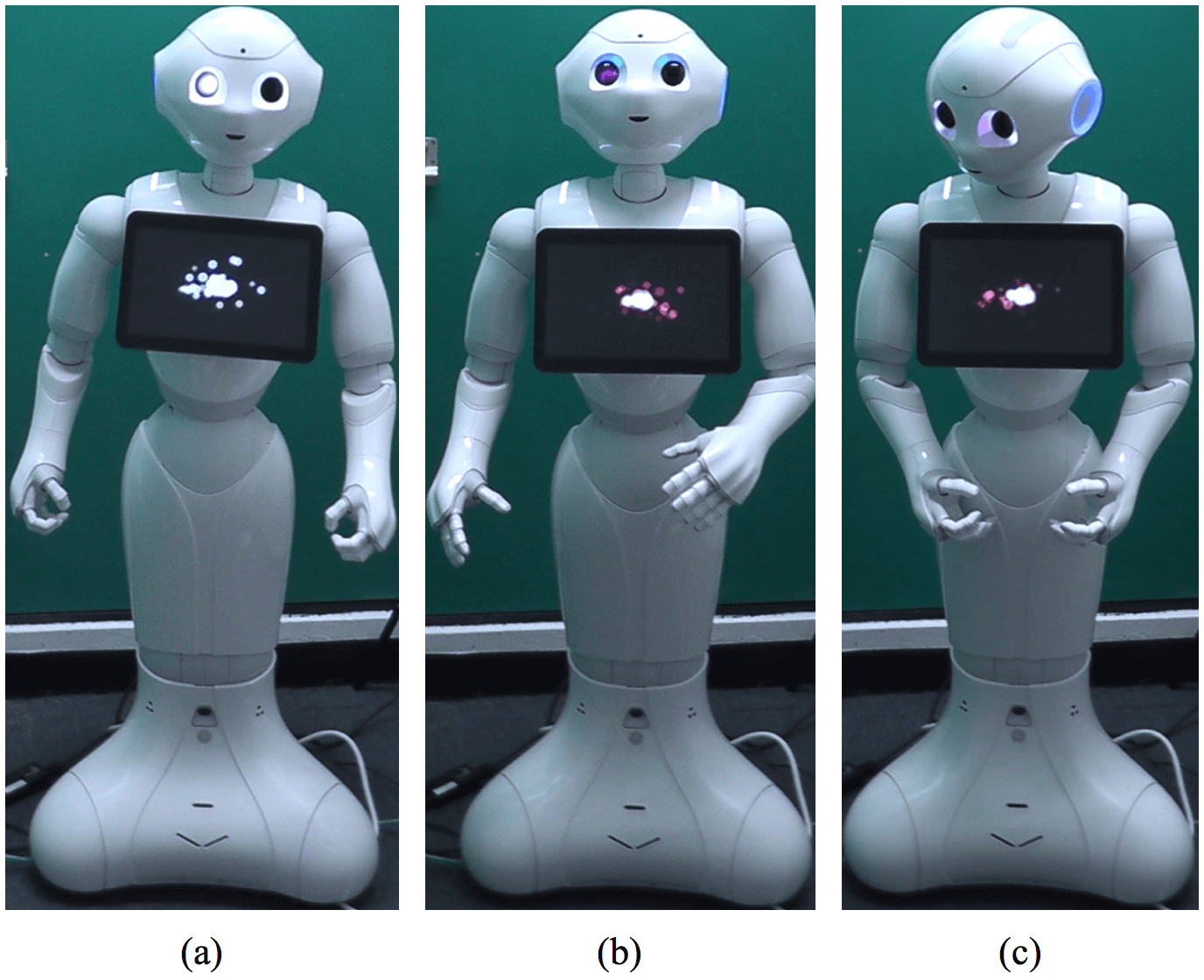}
\caption{Body expressions for Pepper robot when it detects high level of emotional intensities for: (a) joy; (b) surprise; (c) sadness}
\label{fig:body-expr}
\end{figure}

To further test the practical usage, we deploy the proposed method on a physical robot with the task to detect participants' emotional intensities. However, during a real Human-Robot Interaction (HRI), the ground-truths of subjects' emotional intensities are hard to obtain, thus making the evaluation difficult. On the other hand, subjects themselves are well aware of what emotions and intensities they are expressing when interacting the robot. Thus, if the robot can show the estimation results in real-time to the subjects, then the subjects can online evaluate whether the results are accurate or not. Instead of directly showing/telling the detected emotional intensities, we use different robot behaviors to imply the estimates in order to make the interaction more natural and interactive. Specifically, for the easy of robot behavior design, we threshold the emotional intensities into two levels (weak and strong) and only consider three common emotion types (joy, surprise and sadness). Based on this, we design two types of robot behaviors: \emph{none} and \emph{expressive}. In \emph{none}, the robot has no emotional intensity estimation, only reacting with random fillers, e.g., oh and hmm, regardless of subjects' emotions and the intensities. By contrast, in mode \emph{expressive}, the robot will show behaviors based on the detected emotional intensities in real-time. If low intensity is detected, the robot will provide speech feedback, e.g., ``you look so happy!" for joy, ``why are you surprised?" for surprise, and ``what happened to you?" for sadness. If high intensity is detected, the robot will provide gestural feedback, as shown in Figure~\ref{fig:body-expr}. The set of speeches and postures varies for different emotions.

\begin{figure}
\centering
\includegraphics[width=0.9\linewidth]{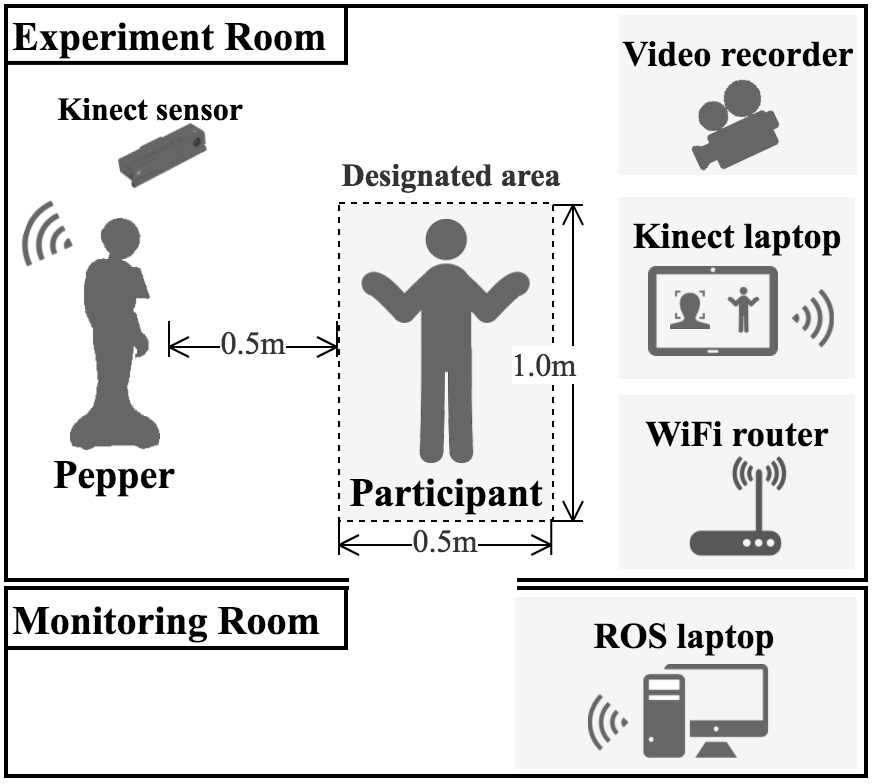}
\caption{Illustration of experiment setup}
\label{fig:exp-setup}
\end{figure}

The field testing was conducted on a Pepper robot with a Kinect sensor, as shown in Figure~\ref{fig:exp-setup}. The subject's body poses were captured by the Kinect Sensor in 25 joints, each with a 3D position in the camera coordinate frame and a 4D rotation quaternion in a local coordinate frame, which are converted to Euler angles based on these formulas\footnote{https://msdn.microsoft.com/en-us/library/hh973073.aspx}. The subjects' faces are also detected and captured by the Kinect sensor, which are then analyzed by an facial analysis tool \footnote{https://azure.microsoft.com/en-us/services/cognitive-services/emotion/} to obtain facial expressions.

We recruit 14 subjects with their informed consents to participate in the field test. In the process, subjects were required to express three emotions (joy, sadness and surprise) via their faces and body poses in two levels of intensities. Subjects were also told about the purpose of the experiment to encourage them to change their emotions and intensities. Each experiment consists of three sessions, and, in each session, the subject was required to perform at least one emotion for around 60s. After each session, subjects need to rate the robot's emotion-perceiving ability in terms of  ``emotion perceiving'' and ``arousal sensing'' on a 7-point Likert scale (1 means very poor and 7 means very good). In addition, they need to rate the robot's other performances derived from \cite{strait2015too,lee2010gracefully}. The whole system was implemented in Robot Operating System (ROS) and runs in real-time with Ubuntu system.

\begin{figure}
\centering
\includegraphics[width=1.0\linewidth]{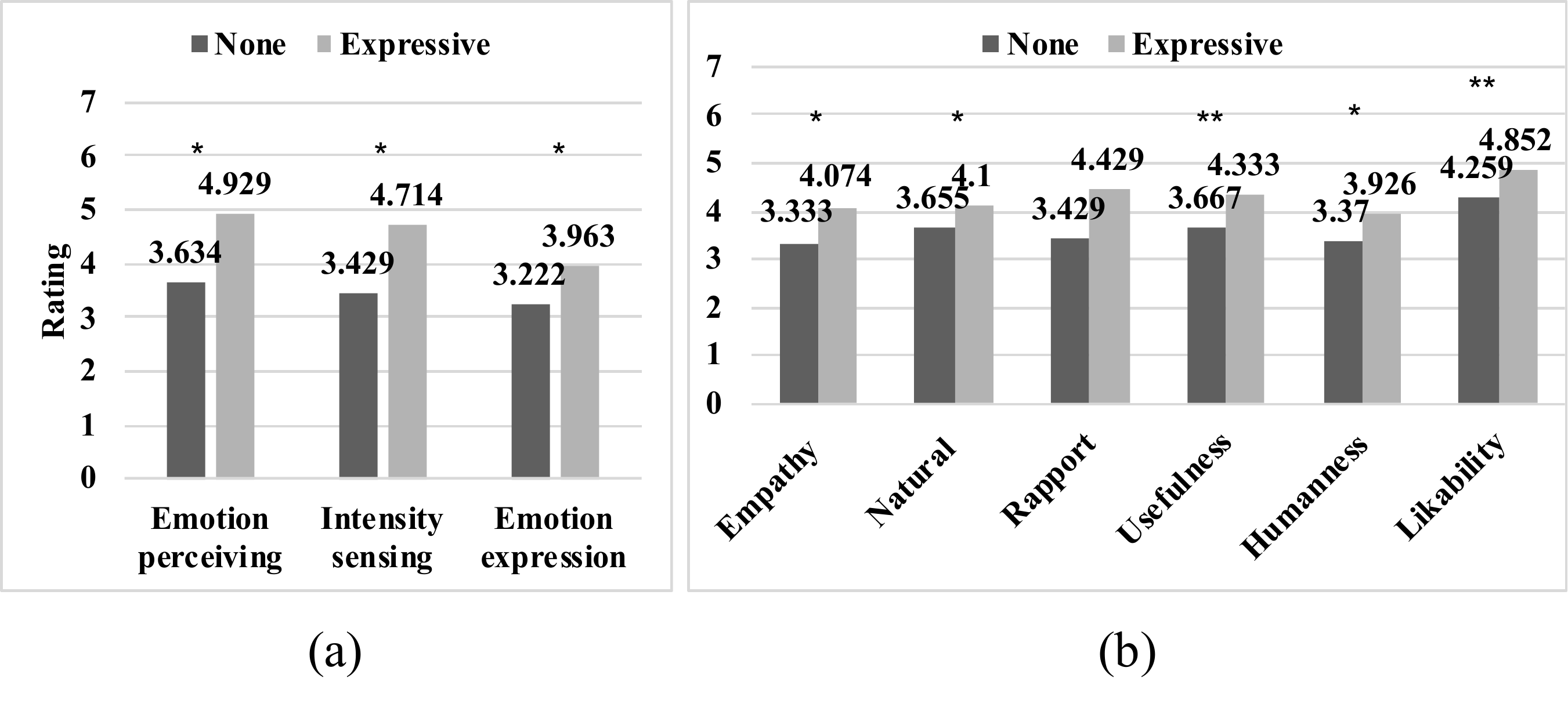}
\caption{Ratings of: (a) robot emotion perceptions(1 means very poor and 7 means very good); (b) robot other performances(1 means very poor and 7 means very good). ($*: p<.05$ and $**: p<.01$)}
\label{fig:hri-experience}
\end{figure}

Ratings are shown in Fig.\ref{fig:hri-experience}. To spot the statistic significance, we use the repeated measures ANOVA to analyze these ratings. First, the manipulation check on Pepper’s expressiveness shows that the manipulation is effective; repeated measures ANOVA, $F (2, 52) = 9.427, p < 0.01, \eta^2 = .266$; Bonferroni post-hoc test $p < 0.01$. The robots using the our method (``expressive'': M = 4.296, SD = 0.183) are indeed perceived to be more expressive than the baseline version (``none'': M = 3.222, SD = 0.252). As for the robot emotion perception, all subjects rated the emotion perceiving, intensity sensing and emotion expression in mode ``expressive'' significantly higher than the correspondences in mode ``none''. Specifically, for emotion perceiving, mode ``expressive'' (M = 4.929,SD = 0.165) significantly exceeds mode none (M = 3.634, SD = 0.372) with repeated measures ANOVA, $F (2, 26) = 4.225, p < 0.05, \eta^2 = .247$ and Bonferroni post-hoc test $p < 0.05$, and, for intensity sensing, mode ``expressive'' (M = 4.714,SD = 0.221) is significantly higher than mode ``none'' (M = 3.429, SD = 0.388) with repeated measures ANOVA, $F (2, 26) = 4.381, p < 0.05, \eta^2 = .252$ and Bonferroni post-hoc test $p < 0.05$. In summary, the proposed method is effective for real-time application in HRI. Furthermore, subjects also gave significantly different ratings on robots' other performances, such as robots' ability to show empathy, to behave naturally, to demonstrate usefulness, to perform human-like and to gain popularity. Overall, ratings  in mode ``expressive'' on various aspects are significantly higher than that in mode ``none''. These results suggest that the proposed method is useful for a real HRI, and the robot equipped with our model can dramatically change subjects' views of its performances.

\section{Conclusion}
We propose a method to estimate emotional intensities from body poses under various affective status. This method adopts the local joint transformations to describe body poses, which are invariant to subjects' body shape differences and their positions against sensors. Moreover, we propose an Long Short-Term Memory Recurrent Neural Network (LSTM-RNN) architecture to model pose descriptors without any handcrafted features. Our quantitative evaluations on a dataset imply that the method accurately predicts emotional intensities with a very high correlation score. The field tests on a physical robot demonstrate that the proposed method can be well applied for practical usage as it enables a humanoid robot (Pepper) to sense subjects' emotional intensities effectively in real-time. Also, subjects reported that the robots with our method are more emotional-sensitive and outperform the robots without in all aspects. 

\section*{Acknowledgements}
This project is sponsored by WeChat-HKUST Joint Laboratory on Artificial Intelligence Technology (WHAT LAB) and Innovation and Technology Fund (ITF) with No. ITS/319/16FP. We also thank WeChat team for their contributions to this paper.



%

\bibliographystyle{IEEEtran}
\bibliography{IEEEabrv,reference}

\end{document}